## Note from the author.

This comment has been originally submitted to *Journal of Applied Physics* on November 4, 2025, following a series of unsatisfactory interactions with the authors and the editorial board regarding several issues identified this paper. The editorial board had initially confirmed that they would send this comment out for review after receiving an answer from authors. After several months without a clear update on the status of my submission, on May 5, 2026 a Retraction Notice has been posted for the original article. The retraction notice does not make any mention of this submitted Comment and, in fact, it implies that the retraction was initiated by the authors, without clarifying who performed the "post-publication review".

In order to provide the academic community with additional details on this matter (and also to show the actual amount of work that had been necessary from my side to convince the authors and the editorial board of the substance of my comment), I have decided to post the submitted Comment on Arxiv.

# Comment on "Angle insensitive filters based on Fabry–Pérot resonance structures" [J. Appl. Phys. 136, 193102 (2024)]

Michele Cotrufo[1*]

[1]The Institute of Optics, University of Rochester, Rochester, New York 14627, USA *michele.cotrufo@rochester.edu

In a recent paper,[1] Cao et al. investigated numerically a structure made of two cascaded metasurfaces to realize angle-independent filters. The proposed device is sketched in **Fig. 1a** (adapted from Fig. 1a of [1]). In Fig. 7a of [1] (reproduced here in **Fig. 1b**), Cao et al. show the numerically calculated transmission of the device versus wavelength and polar angles, in the ranges [1120 nm, 1190 nm] and $[0°, 70°]$. **Fig. 1c** of this Comment reproduces Fig. 7b of [1], and it shows two transmission spectra at $\theta = 0°, 30°$. The results in Figs. 7(a-b) of [1] form the main claim of the manuscript *("Angle insensitive filters")*. Indeed, based on these calculations, the device behavior appears remarkable: a narrow band-pass response, with a linewidth of ~5 nm, is maintained identical (at the same central wavelength) for any angle up to $70°$, with only a minor decrease of the peak transmission. However, despite several attempts and interactions with the authors, we have been unable to replicate these simulations. This Comment is divided into three sections. First, we discuss our attempts at reproducing the authors' calculations. Second, we point out the lack of any physical explanation for such angle-independent behavior. Finally, we point out that the results of Cao et al. can, in fact, be well reproduced by simulations which use **incorrect boundary conditions**.

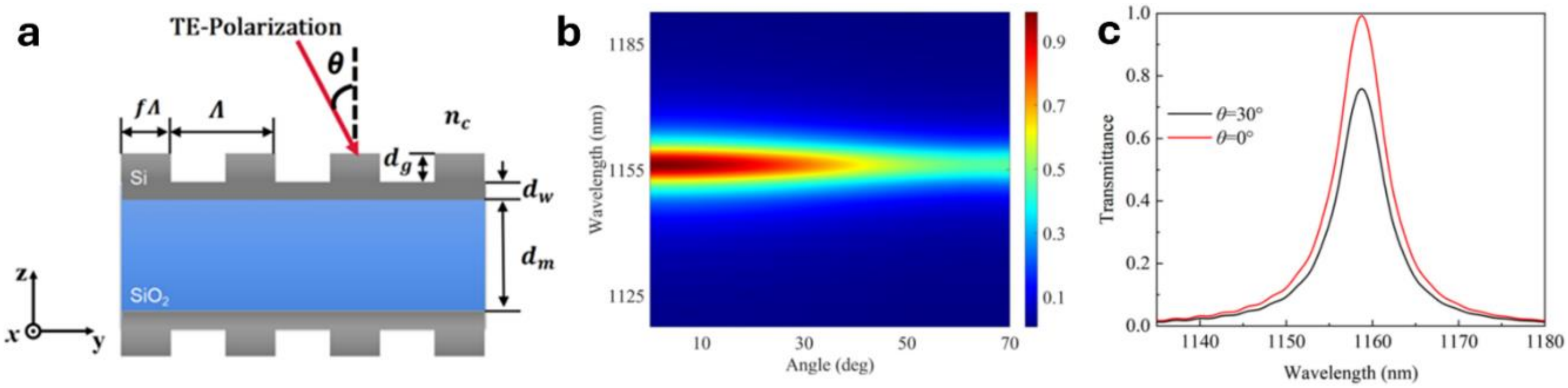

**Figure 1. (a)** Schematic of the structure proposed in [1] (Figure reproduced from Fig. 1a of [1]). **(b-c)** Reproduced from Fig. 7 of [1].

**1. Inability to reproduce the numerical simulations.** In their paper[1], Cao et al. do not disclose the simulation software used and the exact parameters used for the calculations of Fig. 7 (**Figs. 1(b-c)** in this Comment). We have contacted the authors (via the editorial office) to obtain these information; in their response, the authors have clarified[2,3] that the software was FDTD Lumerical (v2020), and that the geometrical parameters in their Fig. 7 were $d_g$=240 nm, $d_w$=240 nm, $d_m$=310 nm, $f$=0.6, $\Lambda$=640 nm. Moreover, the authors have provided the raw data of the plots shown in Fig. 7 of their article. We have used these parameters to attempt to replicate the results by Cao et al. by using three different electromagnetic simulation packages: (1) The same FDTD-based software used by the authors (Lumerical), (2) a finite-element-method (FEM) solver (Comsol), and (3) a Matlab library which implements RCWA (Reticolo[4]). The angle- and wavelength-dependent transmission maps calculated with these three methods [**Figs. 2(a-c)**] agree very well with each other (apart for minor spectral shifts of few nanometers), but they strongly disagree with the data provided by Cao et al. For better comparison, we have plotted the raw data provided by the authors in **Fig. 2d** using the same ranges and colormap as **Figs. 2(a-c)**. Both our calculations and the calculations by Cao et al. predict the presence of a transmission peak close to 1150 nm. In the calculations of Cao et al., the spectral position of this peak remains constant for any angle up to $\theta = 70^o$. Instead, in our calculations the peak shifts substantially as soon as $\theta > 5^o$. For $\theta > 20^o$ the peak has completely shifted away, and other peaks cross the spectrum in the same range. We note that an angular tolerance of $\sim 5°$ is fairly common for a single resonant waveguide grating (see, e.g., Fig. 4 of [5]), without any need of a bilayer cavity arrangement. Moreover, our simulations show that additional narrow peaks appear close to 1160 nm when $\theta > 0$. The appearance of such sharp peaks is generally expected in these grating-like structures for off-normal excitations. We have contacted the authors to help us identify possible reasons for such large discrepancy and we have also requested access to their simulation files to compare them with ours. **The authors have declined to share their files citing intellectual property issues[2].** We have also analyzed the raw data provided by the authors. **Fig. 2e** shows vertical 1D slices from **Fig. 2d**, that is, transmission versus wavelength at select angles, with each curve normalized to their maximum. We note that all curves overlap exactly, including weak oscillations and noise on the tails of the peak. Such a perfect overlap between the transmission lineshapes at largely different angles is extremely unusual for periodically patterned devices, and we are not aware of any example in literature where such a perfect correlation would occur. Additionally, we have found that in the results of Cao et al. (**Fig. 2d**) the dependence of the peak intensity (at $\lambda_{max} = 1152$ nm) versus $\theta$ can be fitted perfectly ($R^2 = 0.99996$) with a rescaled cosine function.

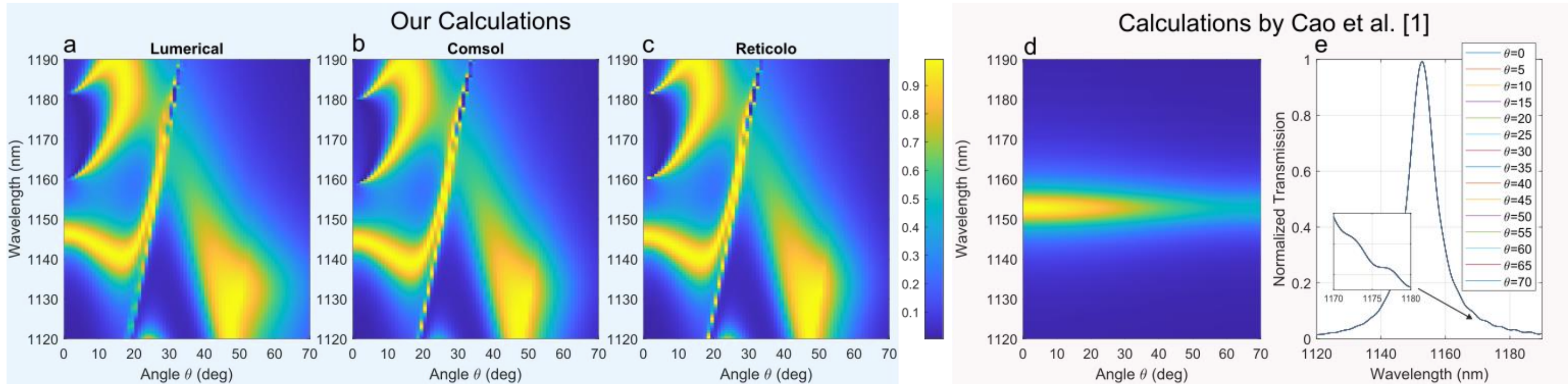

**Figure 2. (a-c)** Results of our numerical calculations performed with (a) Lumerical, (b) Comsol, (c) Reticolo. **(d)** Numerical calculations presented by the authors in Fig. 7 of [1]. The raw data sent to us by the authors has been plotted with the same range as panels (a-c). All color plots in this figure share the same color bar. **(e)** Vertical cross sections from panel d at select angles. Each plot has been normalized to its maximum.

**2. Lack of explanation for the claimed angle-independent behavior, and inconsistency in geometrical parameters between Fig. 7 and Fig. 8.** We also notice that the paper of Cao et al.[1] provides no theoretical explanation, either quantitative or qualitative, for the claimed angle-independent behavior. Instead, most of the paper focuses on (i) general theory of Fabry-Perot cavities (Eqs. 1-5), (ii) considerations on how each grating can be described by an effective refractive index (Eqs. 6-7, Fig. 4), (iii) impact of fabrication tolerance on the normal-incidence transmission spectrum (Fig. 5), dependence of the normal-incidence transmission spectrum on the grating period (Fig. 6). Then, Figure 7 is presented. Commenting Fig. 7 and Fig. 8, the authors state[1]

*"The standing wave modes in the Fabry–Pérot cavity play a crucial role in its angle insensitivity. Clear standing wave patterns can be observed in the electric field diagrams at different incident angles in Fig. 8, and the positions of the standing wave nodes remain almost unchanged as the angle varies. This indicates that the coherent interference structure inside the cavity is not sensitive to changes in the incident angle. Moreover, the entire structure is made of all-dielectric materials, which have lower losses compared to metallic materials, preventing significant attenuation of the standing waves due to material absorption. This results in consistent transmission spectra over a wide range of angles."*

However, this does not provide any "causal" explanation. Rather, it now shifts the question from "why is the transmission spectrum angle-independent" to "why is the standing-wave pattern inside the cavity angle-independent?". The two features (angle-independent transmission spectrum and angle-independent field pattern) are connected, but no explanation is provided for why either of them should occur. Moreover, and importantly, we point out that **the field profiles shown in Fig. 8 of [1] cannot possibly refer to the same structure considered in Fig. 7**. The parameters communicated by the authors[2,3] for their simulations in their Fig. 7 are $d_g$=240 nm, $d_w$=240 nm, $d_m$=310 nm. Thus, the silica layer ($d_m$) should be thinner than the overall grating thickness ($d_g + d_w$). Instead, Fig. 8 of [1] clearly shows that the silica layer is much thicker than each of the two metasurfaces. Thus, either Fig. 8 was not calculated for the same structure as Fig. 7, or the parameters disclosed by the authors for Fig. 7 are wrong.

**3. Exploring possible reasons for the discrepancy between the author's simulations and ours.** In a subsequent communication[6], the authors have suggested that the discrepancy between our results and theirs "*are likely due to variations in simulation settings, including boundary conditions, mesh resolution, and incident beam parameters.*". We agree that small numerical variations can occur in electromagnetic simulations due to different mesh resolutions and differences between the numerical methods used (e.g. FDTD vs FEM). Examples of such small deviations are indeed visible also in our own simulations in **Figs. 2(a-c)**, and they result in small spectral shifts of the peaks and dips. However, we argue that the strongly different behavior between our results [**Figs. 2(a-c)**] and theirs (**Fig. 2d)** cannot be realistically attributed to simulation accuracy. Nonetheless, to address this possibility, we have conducted a series of systematic simulation campaigns. We have: (i) considered different simulation accuracies, obtained by increasing the mesh resolution in Comsol and the number of Fourier harmonics in the RCWA simulations; (ii) swept the values of almost all geometrical parameters around the nominal ones, and (iii) considered any possible combination of impinging polarization and azimuthal angle (to account for the possibility that the excitation geometry in Fig. 1a does not match the actual one). The results of these simulations are reported in the Supplemental Material, and they all show a consistent and large deviation from the results in Fig. 7 of [1]. In particular, none of these additional calculations show the angle-independent behavior claimed by Cao et al. in [1].

As mentioned in the previous paragraph, the authors also suggested[6] that that the discrepancy between our results and theirs could be due to variations in the *"boundary conditions"*. We are unsure how to interpret this suggestion. Boundary conditions (BCs) are set by the problem under study, and they cannot be arbitrarily chosen. To simulate the injection of a tilted wave in a periodic structure, Bloch/Floquet (B/F) BCs must be applied on the lateral edges of the domain. Assuming that the structure is periodic along the y axis and that the excitation wave-vector is $\boldsymbol{k} = [0, k_y, k_z]$, the B/F BCs dictate that $\boldsymbol{E}(x, y + \Lambda, z) = \boldsymbol{E}(x, y, z)e^{ik_y\Lambda}$. Using other BCs (such as 'periodicity', i.e., $\boldsymbol{E}(x, y + \Lambda, z) = \boldsymbol{E}(x, y, z)$) will not correctly inject a tilted excitation. "Inspired" by this comment from the authors, we have repeated our Lumerical simulations by purposedly using wrong BCs. In particular, we have repeated the simulations shown in **Fig. 2a** of this Comment, but by setting the BCs to periodicity instead of B/F, while keeping an excitation port with a finite tilted angle $\theta$ (additional details are available in the Supplemental Material). **Figure 3a** shows the numerical results obtained with this approach. Apart from a small spectral shift, an excellent match with the results shown by Cao et al. is obtained. This is further shown in Figs. **3(b-c)**, where we compare our calculations (with wrong BCs) to the ones of Cao et al. at three select angles. Our calculations with wrong BCs reproduce very well both the (apparent) angle-insensitivity and the reduction in peak transmission. We emphasize again that in these simulations we purposedly set wrong BCs (periodicity instead of B/F), and therefore the obtained results cannot be considered as correct. In particular, as discussed in the Supplemental Material, using such incorrect BCs leads that a wave being injected always at normal incidence, even if the port is set up to inject a tilted excitation. The only effect of setting $\theta > 0$ in the injection port is that the amplitude of the injected wave gets artificially lowered, which can very-well explain why in the data by Cao et al. the peak transmission decreases following a cosine function (see Figure S4 in Supplemental Material).

In conclusion, based on our analysis we believe that the discrepancies between the results by Cao et al. and our calculations can be explained only in two ways: either **(i)** the actual geometrical and/or material parameters used by Cao et al. for the calculations in Fig. 7 of [1] are largely different from the ones that have been reported in the paper and/or communicated to us, or **(ii)** the numerical simulations performed by Cao et al. are not correct. In section 3 and **Fig. 3** of this Comment we have shown that, for example, using incorrect boundary conditions reproduces extremely well their data (but of course, the results are unphysical). The only way to conclusively clarify this is for the authors to share their simulation files. As mentioned above, the authors have declined to do so citing intellectual property issues[2]. Given that the authors have already disclosed the parameters of their structure and published a paper about it, it is unclear why sharing the simulations files would impact any intellectual property claim. All codes and scripts required to reproduce our calculations are available in an online repository[7].

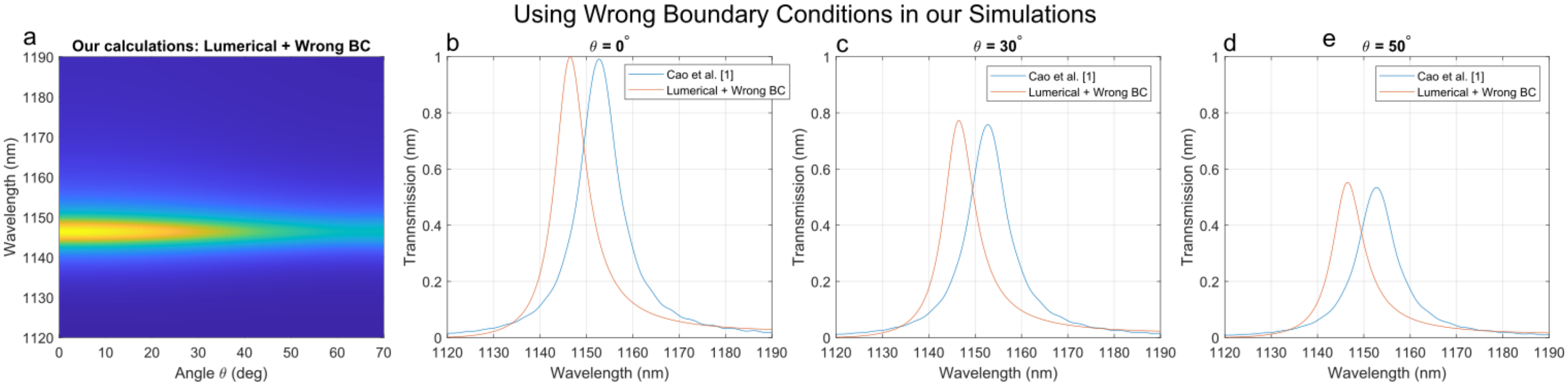

**Figure 3. (a)** Results of our numerical calculations performed with Lumerical and using wrong boundary conditions (periodicity instead of Bloch/Floquet). **(b-c)** Comparison between our simulations with wrong boundary conditions (panel a) and the results of Cao et al., at three angles.

## Supplemental Material

The Supplemental Material, including high-resolution version of the pictures, is available in the Zenodo repository associated with this Comment, **https://doi.org/10.5281/zenodo.17488428**

# References

1. Cao, S., Chen, N. & Jiang, Y. Angle insensitive filters based on Fabry–Pérot resonance structures. *Journal of Applied Physics* **136**, (2024).
2. Private Communication from Authors via the Editorial Office, received by us on July 3, 2025.
3. Private Communication from Authors via the Editorial Office, received by us on July 7, 2025.
4. Hugonin, J. P. & Lalanne, P. Reticolo software for grating analysis. *arXiv preprint arXiv:2101.00901* (2021).
5. Markowitz, M. *et al.* Tailored resonant waveguide gratings for augmented reality. *Optics Express* **30**, 20469–20481 (2022).
6. Private Communication from Authors via the Editorial Office, received by us on October 17, 2025.
7. Cotrufo, M. Code and models to reproduce the calculations in ‘Comment on “Angle insensitive filters based on Fabry–Pérot resonance structures” [J. Appl. Phys. 136, 193102 (2024)]’. **https://doi.org/10.5281/zenodo.17488428**